\documentclass{elsart}
\journal{New Astronomy}
\usepackage{natbib}

\usepackage{epsfig}

\usepackage{amssymb}

\def\apj{Astrophys. J.}
\def\apjs{Astrophys. J. Supp.}
\def\mnras{Mon. Not. R. Astron. Soc.}

\def\ghz{\rm GHz}
\def\lsim{\mathrel{\rlap{\lower4pt\hbox{\hskip1pt$\sim$}}
    \raise1pt\hbox{$<$}}}                % less than or approx. symbol
\def\gsim{\mathrel{\rlap{\lower4pt\hbox{\hskip1pt$\sim$}}
    \raise1pt\hbox{$>$}}}       

\begin{document}

\begin{frontmatter}



\title{Angular power spectrum of CMB anisotropy from WMAP}


\author{{\bf Tarun Souradeep}$^1$, Rajib Saha$^{1,2}$ and Pankaj
Jain$^2$} \address{$^1$Inter-University Centre for Astronomy and
Astrophysics (IUCAA),\\ Post Bag 4, Ganeshkhind, Pune 411~007,
India.\\ E-mail: tarun@iucaa.ernet.in;rajib@iucaa.ernet.in
\\$^2$Physics Department, Indian Institute of Technology, Kanpur, U.P,
208016, India. E-mail:rajib@iitk.ac.in;pkjain@iitk.ac.in}
\begin{abstract}
% Text of abstractve 

The remarkable improvement in the estimates of different cosmological
parameters in recent years has been largely spearheaded by accurate
measurements of the angular power spectrum of Cosmic Microwave
Background (CMB) radiation.  This has required removal of foreground
contamination as well as detector noise bias with reliability and
precision. Recently, a novel {\em model-independent} method for the
estimation of CMB angular power spectrum from multi-frequency
observations has been proposed and implemented on the first year WMAP
(WMAP-1) data by Saha et al.~2006. We review the results from WMAP-1
and also present the new angular power spectrum based on three years
of the WMAP data (WMAP-3). Previous estimates have depended on
foreground templates built using extraneous observational input to
remove foreground contamination.  {\em This is the first demonstration
that the CMB angular spectrum can be reliably estimated with precision
from a self contained analysis of the WMAP data}.  The primary product
of WMAP are the observations of CMB in $10$ independent difference
assemblies (DA) distributed over $5$ frequency bands that have
uncorrelated noise.  Our method utilizes maximum information available
within WMAP data by linearly combining DA maps from different
frequencies to remove foregrounds and estimating the power spectrum
from the $24$ cross power spectra of clean maps that have independent
noise.  An important merit of the method is that the expected residual
power from unresolved point sources is significantly tempered to a
constant offset at large multipoles (in contrast to the $\sim l^2$
contribution expected from a Poisson distribution) leading to a small
correction at large multipoles. Hence, the power spectrum estimates
are less susceptible to uncertainties in the model of point sources.
\end{abstract}


\end{frontmatter}

% main text
\section{Introduction}

Remarkable progress in cosmology has been made due to the measurements
of the anisotropy in the cosmic microwave background (CMB) over the
past decade.  The extraction of the angular power spectrum of the CMB
anisotropy is complicated by foreground emission within our galaxy and
extragalactic radio sources, as well, as the detector
noise~\cite{bouc_gisp99,Tegmark96}. It is established that the CMB
follows a blackbody distribution to high accuracy~\cite{FIRAS}.
Hence, foreground emissions may be removed by exploiting the fact that
their contributions in different spectral bands are considerably
different while the CMB power spectrum is same in all the
bands~\cite{Dodelson, Tegmark2000a, Bennett1,Tegmark98}.  Different
approaches to foreground removal have been proposed in the
literature~\cite{bouc_gisp99,Tegmark96,hob98,main0203,erik06}.

The Wilkinson Microwave Anisotropy Probe (WMAP) observes in $5$
frequency bands at $23~\ghz$~(K), $33~\ghz$~(Ka), $41~\ghz$~(Q),
$61~\ghz$~(V) and $94~\ghz$~(W).  In the first data release, the WMAP
team removed the galactic foreground signal using a template fitting
method based on a model of synchrotron, free free and dust emission in
our galaxy \cite{Bennett1}. The sky map around the galactic plane and
around known extragalactic point sources were masked out and the CMB
power spectrum was then obtained from cross power spectra of
independent difference assemblies in the $41~\ghz$, $61~\ghz$ and
$94~\ghz$ foreground cleaned maps~\cite{hin_wmap03}.

A model independent removal of foregrounds has been proposed in the
literature~\cite{Tegmark96}.  The method has also been implemented on
the WMAP data in order to create a foreground cleaned
map~\cite{Tegmark}. The main advantage of this method is that it does
not make any additional assumptions regarding the nature of the
foregrounds.  Furthermore, the procedure is computationally fast.  The
foreground emissions are removed by combining the five different WMAP
bands by weights which depend both on the angular scale and on the
location in the sky (divided into regions based on `cleanliness').
However, the power spectrum recovered from the single foreground
cleaned map has severe excess power at large multipole moments due to
amplification of detector noise bias in the autocorrelation power
spectrum beyond the beam resolution.

The prime objective of our method~\cite{sah06} is to adapt and extend
the foreground cleaning approach to power spectrum estimation by
retaining the ability to remove detector noise bias exploiting the
fact that it is uncorrelated among the different Difference Assemblies
(DA) \cite{hin_wmap03,Jarosik}. The WMAP data uses 10 DA's \cite{Bennett,
Bennett2, WMAP, Hinshaw1}, one each for K and Ka bands, two for Q
band, two for V band and four for W band.  We label these as K, Ka,
Q1, Q2,V1, V2, W1, W2, W3, and W4 respectively. We eliminate the
detector noise bias using cross power spectra and provide a model
independent extraction of CMB power spectrum from WMAP first year
data.  So far, only the three highest frequency channels observed by
WMAP have been used to extract CMB power spectrum and the foreground
removal has used foreground templates based on extrapolated flux from
measurements at frequencies far removed from observational frequencies
of WMAP~\cite{hin_wmap03,Pablo,Patanchon}.  We present a more general
procedure where we use observations from all the five frequency
channels of WMAP and do not use any observational input extraneous to
the WMAP data set.

\section{Methodology}

The entire method can be neatly split into three distinct sections --
\begin{itemize}
\item{} foreground cleaning, 

\item{} power spectrum estimation, and, 

\item{} correcting for known systematics. 
\end{itemize}

We discuss in some detail the
systematic correction for residual power due to unresolved sources
\footnote{Study of other effects, such as beam non-circularity is in
progress~\cite{mit04,irv_noncirc}.}.  A merit of our approach is that
the expected residual power in the cross spectra from combined maps
has a flat dependence with increasing multipoles -- a strongly
tempered dependence given that the contribution in individual DA maps
scales as $l^2$ . (This feature holds promise for experiments at
higher angular resolution.)

The three parts of the current methodology are logically modular and
it is possible to modify and improve each one relatively independent
of the other. Specifically, we have employed the Pseudo-$C_l$
(MASTER~\cite{Hivon}) power spectrum estimation from the foreground
cleaned maps which is fast but sub-optimal for the low $l$ (in
practice). More appropriate approaches have been
proposed~\cite{knox01,jew04,wan04}. We are exploring
improvements on this front.

\subsection{Foreground Cleaning}

The foreground cleaning stage of our method is adapted from Tegmark \&
Efstathiou (1996) and Tegmark {\it et al.} (2003). In Tegmark {\it et
al.} (2003), a foreground cleaned map is obtained by linearly
combining 5 maps corresponding to one each for the different WMAP
frequency channels. For the Q, V and W frequency channels, where more
than one maps were available, an averaged map was used. However,
averaging over the DA maps in a given frequency channel precludes any
possibility of removing detector noise bias using cross
correlation. In our method we linearly combine maps corresponding to a
set of 4 DA maps at different frequencies. We treat K and Ka maps
effectively as the observation of CMB in two different DA (at the
lowest frequencies). Therefore we use K and Ka maps in separate
combinations. In case of W band 4 DA maps are available. We simply
form an averaged map taking two of them at a time and form effectively
6 DA maps corresponding to W band.  W$ij$ represents simply an
averaged map obtained from the $i^{\rm th}$ and $j^{\rm th}$ DA of W
band~\footnote{Other variations are possible and some explored. We
defer a discussion to a more detailed publication~\cite{Saha}.} .  In
table \ref{tab:combination} we list all the $48$ possible {linear
combinations of the DA maps that lead to `cleaned' maps, C${\bf i}$
and CA${\bf i}$'s, where {${\bf i}$ = 1, 2, \ldots, 24}.

\begin{table}
\centering
\scriptsize
\begin{tabular}{|l |l|}
\hline 
                         &                           \\
(K,KA)+Q1+V1+W12=(C1,CA1)&(K,KA)+Q1+V2+W12=(C13,CA13)\\
% \hline                                                                      
(K,KA)+Q1+V1+W13=(C2,CA2)&(K,KA)+Q1+V2+W13=(C14,CA14)\\
%    \hline                                                                     
(K,KA)+Q1+V1+W14=(C3,CA3)&(K,KA)+Q1+V2+W14=(C15,CA15)\\
%    \hline                                                                     
(K,KA)+Q1+V1+W23=(C4,CA4)&(K,KA)+Q1+V2+W23=(C16,CA16)\\
%   \hline                                                                     
(K,KA)+Q1+V1+W24=(C5,CA5)&(K,KA)+Q1+V2+W24=(C17,CA17)\\
%   \hline                                                                 
(K,KA)+Q1+V1+W34=(C6,CA6)&(K,KA)+Q1+V2+W34=(C18,CA18)\\
%   \hline                                                                 
(K,KA)+Q2+V2+W12=(C7,CA7)&(K,KA)+Q2+V1+W12=(C19,CA19)\\
%    \hline                                                                   
(K,KA)+Q2+V2+W13=(C8,CA8)&(K,KA)+Q2+V1+W13=(C20,CA20)\\
%   \hline                                                                   
(K,KA)+Q2+V2+W14=(C9,CA9)&(K,KA)+Q2+V1+W14=(C21,CA21)\\ 
%   \hline                                                                   
(K,KA)+Q2+V2+W23=(C10,CA10)&(K,KA)+Q2+V1+W23=(C22,CA22)\\ 
%   \hline                                                                   
(K,KA)+Q2+V2+W24=(C11,CA11)&(K,KA)+Q2+V1+W24=(C23,CA23)\\ 
%   \hline                                                                    
(K,KA)+Q2+V2+W34=(C12,CA12)&(K,KA)+Q2+V1+W34=(C24,CA24)\\
                           &                           \\
 \hline 
\end{tabular} 
 \caption{ List of the 48 possible combinations of the DA maps that
lead to 48 cleaned maps (C{\bf i}, CA{\bf i}), {\bf i}=1,2,..,24..}
\label{tab:combination}
\end{table}

The goal is to determine a set of optimal weights for each multipole,
\begin{equation}
{{\bf [W_l]}=(w_l^1, w_l^2, w_l^3, w_l^4)}
\end{equation} 
that specifies the linear combination of $4$ DA in the combination,
which lead to a `cleaned' map with minimal power

\begin{equation}
a_{lm}^{\rm Clean}=\sum_{i=1}^{i=4}w_l^{i}\frac{a_{lm}^i}{B_l^i}\,,
\label{c_map}
\end{equation}
where $a_{lm}^i$ is spherical harmonic transform of map and $B^i_l$ is
the appropriate beam function for the channel $i$.  The condition that
the achromatic CMB signal remains untouched during cleaning is encoded
as the constraint
\begin{equation}
\bf [W_l][e]=[e]^T[W]^T=1 \,,
\label{constraint}
\end{equation}
where ${\bf [e] }=(1,1,1,1)^T$ is a $4\times 1$ column vector with
unit elements.  The constraint implies that the total power in the
cleaned map
\begin{equation}
C_l^{Clean}={\bf [W_l][C_l][W_l]^T}=C_l^{S}+\bf [W_l][C_l^{(R)}]\bf[W_l]^T
\label{cleaned power}
\end{equation}
where $C_l^{S}$ is the CMB power spectrum and the last term is the
unwanted (non-cosmic, frequency dependent) contribution from the
foreground and other contaminants.

The optimum weights to combine $4$ different frequency channels such
as to minimize the second term on the {\it rhs} of eq.~\ref{cleaned
power} subject to the constraint that CMB is untouched
(eq.~\ref{constraint}) is readily obtained as
~\cite{Tegmark96,Tegmark}
\begin{equation}
\bf [W_l]={[e]^T[C_l]^{-1}}/\left({[e]^T[C_l]^{-1}[e]}\right).
\label{weight}
\end{equation}
 Here the matrix
\begin{equation}
{\bf [C_l]}\equiv C^{ij}_l=\frac{1}{2
l+1}\sum_{m=-l}^{m=l} \frac{a^{i}_{lm}a^{j*}_{lm}}{B_l^i B_l^j}\,.
\end{equation}
In practice, we smooth all the elements of the ${\mathbf C_l}$ using a
moving average window over $\Delta l = 11$ before deconvolving by the
beam function.  This avoids the possibility of an occasional singular
${\mathbf C_l}$ that cannot be inverted.  The entire cleaning
procedure is automated and takes approximately $3$ hours on a $16$
alpha processor machine to get the $48$ cleaned maps.
%One of the cleaned maps, C8, is shown in the Fig.~\ref{c_map1}.
In all the $48$ maps some residual foreground contamination is visibly
present along a small narrow strip on the galactic plane. For the
angular power estimation that follows, the Kp2 mask employed suffices
to mask the contaminated region in all the $48$ maps.  In ongoing
work, we assess the quality of foreground cleaning in these maps using
the Bipolar power spectrum
method~\cite{Amir1,Amir2,Amir3,irv_stataniso}.

\subsection{Power Spectrum Estimation}
\label{power_spectrum_estimation}
   
\begin{figure}
\centering\includegraphics[scale=0.5,angle = -90]{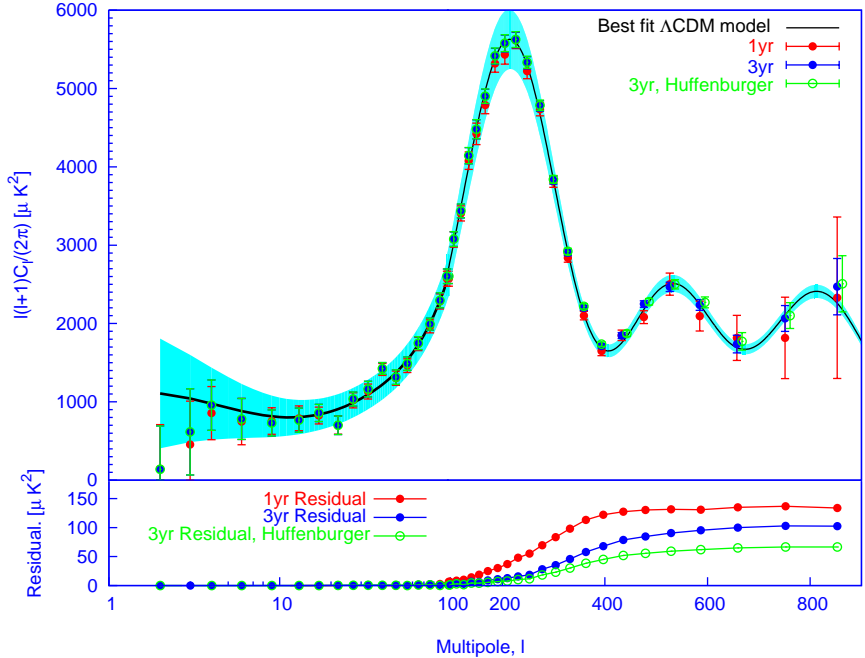}
\caption{The final angular power spectrum from WMAP-1 (red) and WMAP-3
(blue \& green) data are shown in the upper panel. The lower panel
shows the correction made for the residual power from unresolved point
sources based on the point source models published by WMAP team for
the corresponding year. The green curve in the panels correspond to
the WMAP-3 spectrum (upper), and the corresponding correction (lower)
obtained using a revised point source model suggested by Huffenberger
et al. 2006. Note the tempered behavior of the point source correction
in our method (also see fig.~\protect{\ref{ps_plus_5_channel}}). The
multipole range is log scale for $l<100$, and linear, thereafter.}
\label{WMAP13PScorr}
\end{figure} 

Independent estimates of the power spectrum can be obtained by cross
correlation power spectra
\begin{equation}
C^{\mathbf{ij}}_l = (2l+1)^{-1} \sum_m a_{lm}^{\mathbf i} (a_{lm}^{\mathbf j})^*  
\end{equation}
of pairs of cleaned maps C{\bf i} \& C{\bf j} chosen such that they
share no common DA.  This is ensures a zero noise bias of the
estimates. A Kp2 mask is applied to the maps to mask out foreground
contaminated regions and accounted for by de-biasing the pseudo-$C_l$
estimate using the coupling (bias) matrix corresponding to the Kp2
mask, appropriate circularized beam transform and pixel
window~\cite{Hivon}.  The left panel of figure \ref{each_bin} plots
and lists the $24$ cross power spectra for which the noise bias is
zero and closely match each other for $l< 540$. (This is in contrast to
pre-point source corrected cross power spectra in WMAP-1 power spectrum
estimation shown in the right panel -- a merit of our method that we
discuss in the next section.)

An `Uniform average' power spectrum is then obtained by combining the
$24$ cross power spectra with equal weights~\footnote{There exists the
additional freedom to choose optimal weights for combining the 24
cross-power spectra.}. The power
spectrum is then corrected for residual power from unresolved point
sources as described in the next subsection. The error bars have been
estimated with comprehensive simulations described in
$\S$\ref{Errorbars}.  The final power spectrum is binned in the same manner as
the WMAP's published result for ease of comparison.

\begin{figure}
\centering\includegraphics[scale=0.27,angle=-90]{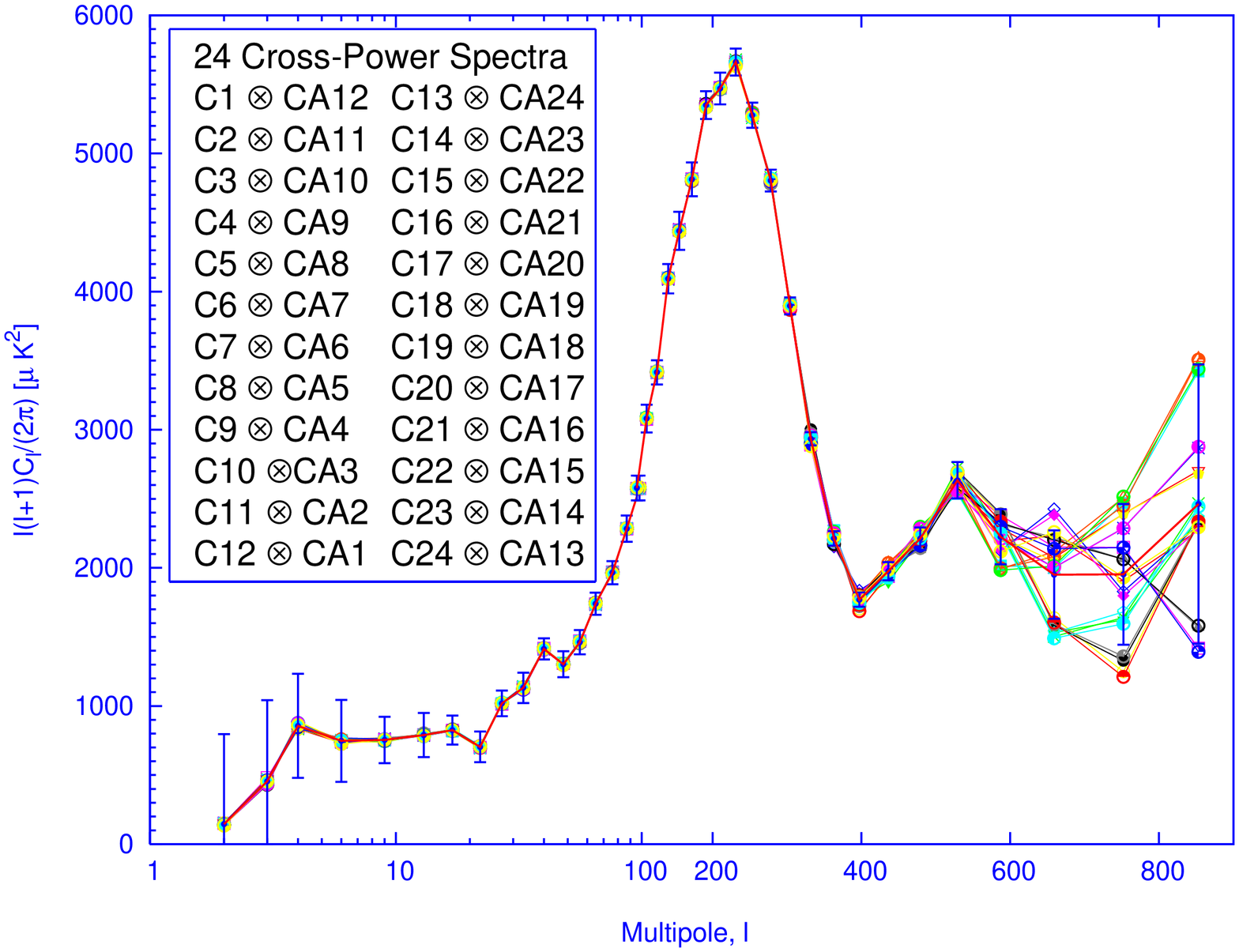}
\includegraphics[scale=0.27,angle=-90]{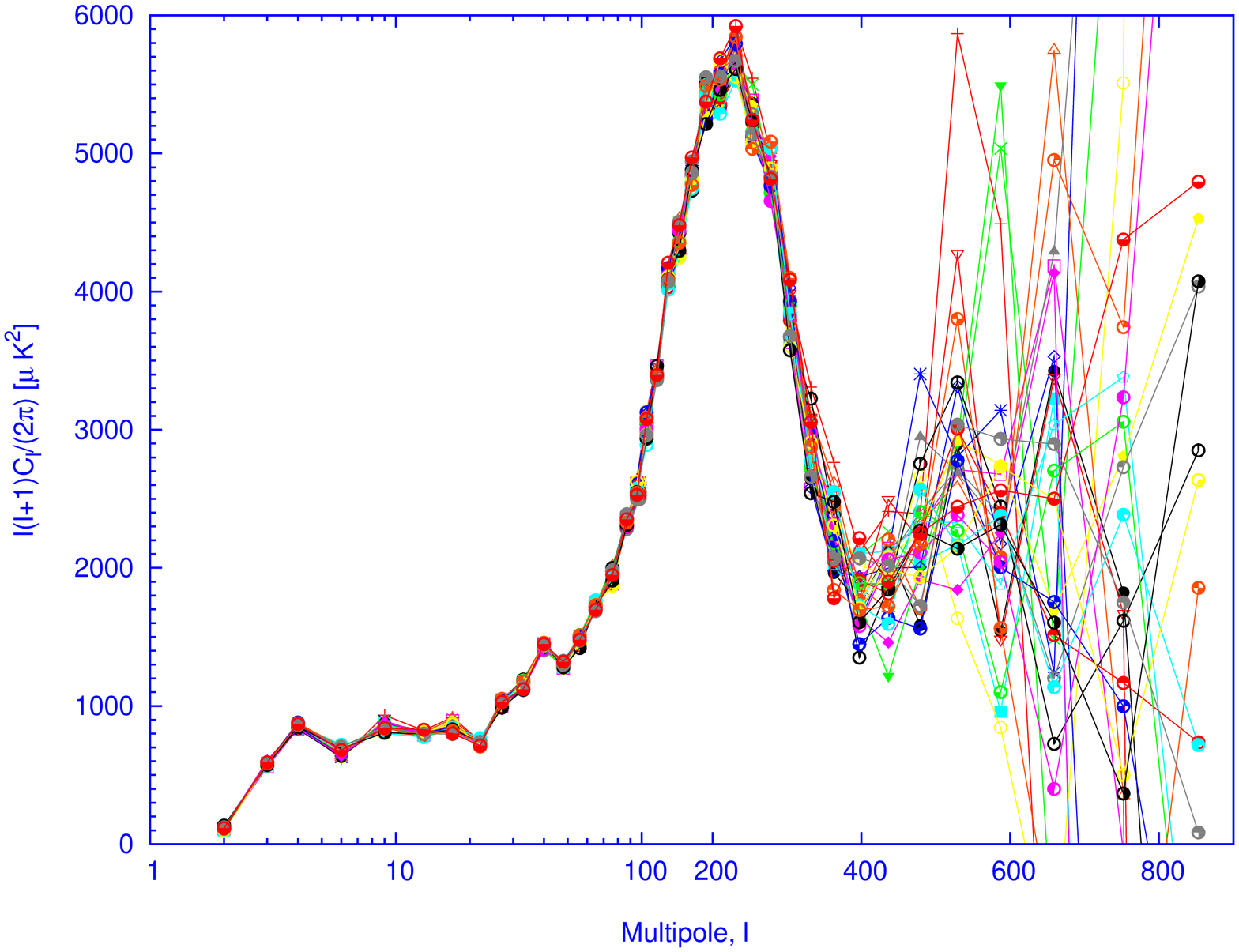}
\caption{{\em Left:} Plot of the $24$ individual cross power spectra
corresponding to the cross correlations listed in this figure are show
very small dispersion for $l\lsim 540$. The average power spectrum is
plotted in red line and blue error bars. {\em Right:} Plot of the $28$
cross-power spectra in the WMAP-1 team analysis prior to point source
corrections show a large scatter at large $l$.  The multipole range is
log scale for $l<100$, and linear, thereafter.}
\label{each_bin}
\end{figure}

The power spectra from our independent analyzes is entirely consistent
with the WMAP-1~\cite{hin_wmap03,sah06} and
WMAP-3~\cite{hin_wmap06,wmap3reanal} team results. We find a
suppression of power in the quadrupole and octopole moments consistent
with WMAP-1 published result. However, in WMAP-1 our quadrupole moment
($ 146 \mu K^2 $) is a little larger than WMAP-1 result($ 123 \mu K^2
$) and Octopole ($ 455 \mu K^2$) is less than WMAP-1 result ($ 611 \mu
K^2$).  Our WMAP-1 spectrum does not show the `bite' like feature
present in WMAP's power spectrum at the first acoustic peak reported
by WMAP-1~\cite{hin_wmap03}. In the three year results, the differences
at low $l\le 10$ between WMAP result and our is primarily due to the
fact that WMAP report the maximum likelihood values for $l\le 10$ (and
we employ pseudo-$C_l$ estimation). At large $l$, our results are
consistent with other estimates as shown in a comprehensive reanalysis
of WMAP-3 carried out jointly with five independent, international CMB
analysis groups (see MASTERint in Ref.~\cite{wmap3reanal}).

A quadratic fit to the peaks and troughs of the binned WMAP-1 {\it
[WMAP-3]} power spectrum shown in the figure~\ref{peak_fit}, locates
the first acoustic peak at $l = 219.8\pm 0.8~~\mathit{[219.9\pm 0.8]} $ with
amplitude $\Delta T_l = 74.1 \pm 0.3 ~~\mathit{[74.4\pm 0.3]} \mu K $ , the
second acoustic peak at $l = 544 \pm 17 ~~\mathit{[539.5\pm 3.8]} $ with
amplitude $\Delta T_l = 48.3 \pm 1.2 ~~\mathit{[49.4\pm 0.4]} \mu K $ and the
first trough at $l = 419.2 \pm 5.6 ~~\mathit{[417.7\pm 3.2]} $ with amplitude
$\Delta T_l = 41.7 \pm 1 ~~\mathit{[41.3\pm 0.6]} \mu K$.

\begin{figure}
\centering
\includegraphics[scale= 0.27,angle = -90]{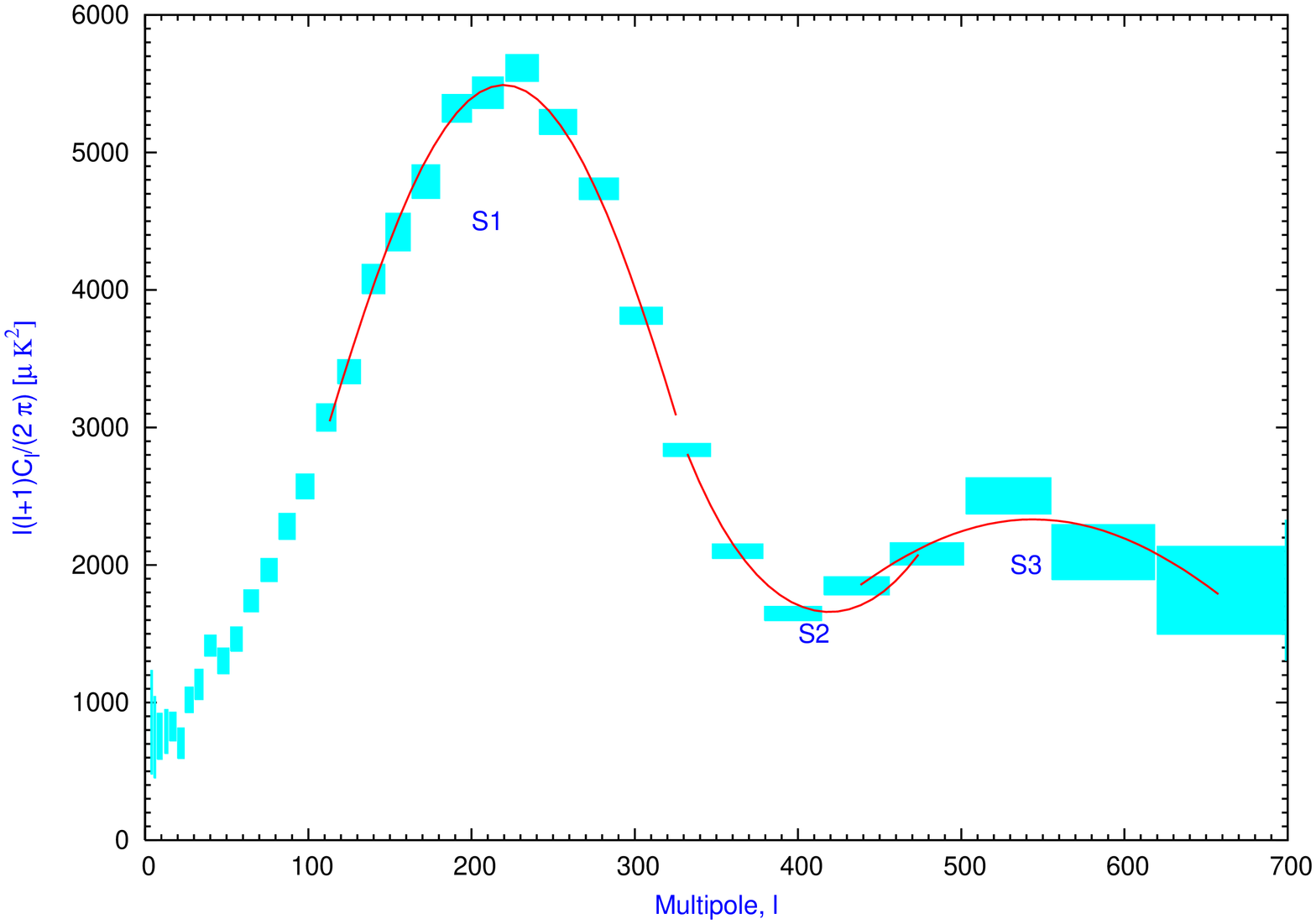}
\includegraphics[scale= 0.27,angle = -90]{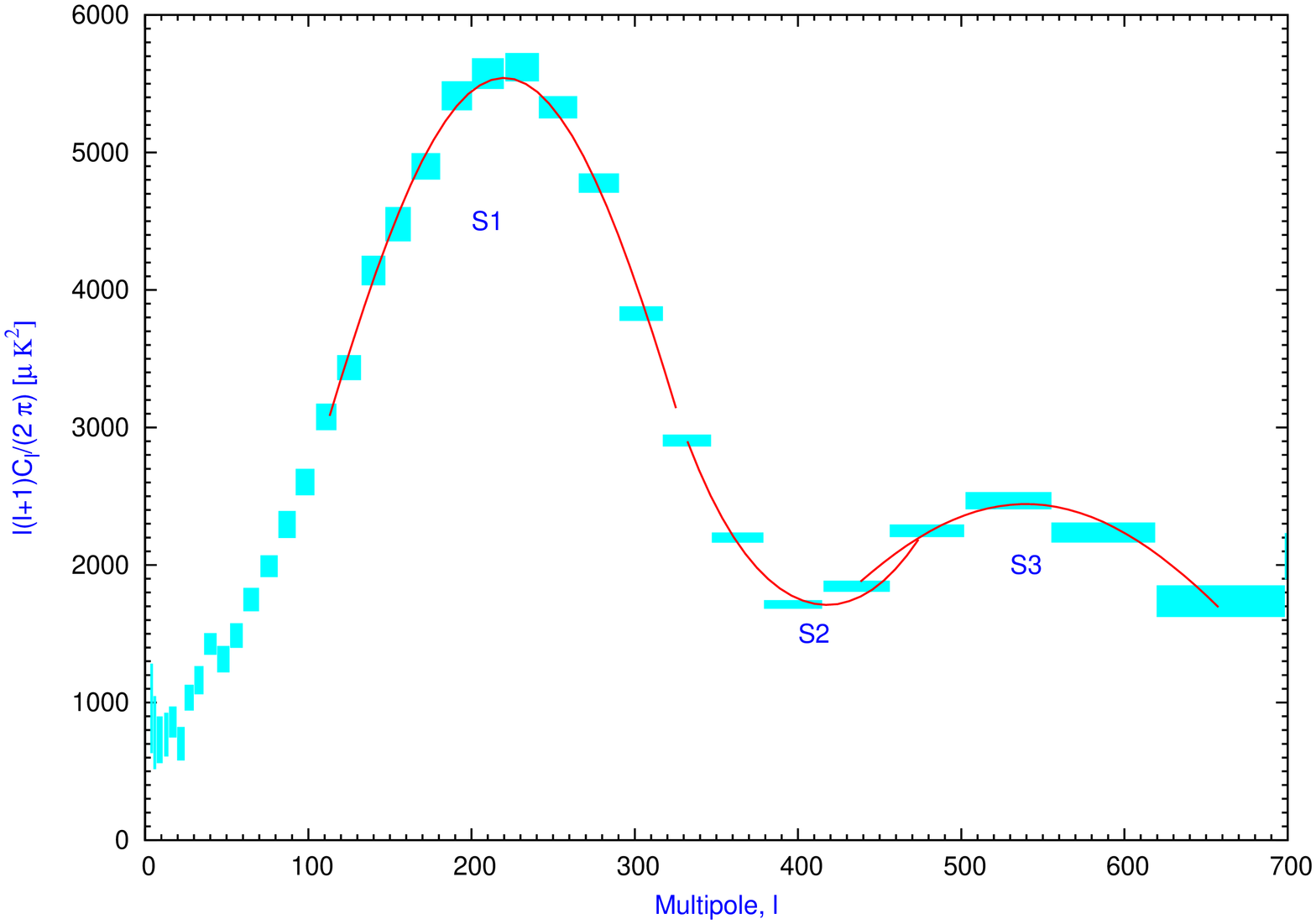}
\caption{ The red line shows the quadratic function fitted to the
peaks and troughs of the final binned angular power spectrum for
WMAP-1 data (left) and WMAP-3 (right). S1, S2 and S3 are the 3
different sections of the spectrum on which fits are performed
individually. The power spectrum estimates with error bars and
multipole band are represented as the shaded boxes.}
\label{peak_fit}
\end{figure}

\subsection{Residual unresolved point source correction}

The residual power contamination in the `Uniform average' power
spectrum from unresolved point sources can be estimated by running
through our analysis the frequency dependent power obtained from the
same source model used by WMAP team to correct for this
contaminant~\cite{hin_wmap03,hin_wmap06}. WMAP has a fitted a simple
model with a constant spectral index over the entire sky for the
residual power in cross-spectra $C^{(PS)ij}_l$ of frequency channel
$i$ and $j$ from unresolved point sources as~\cite{Bennett1}
\begin{equation}
C^{(PS)ij}_l =
A\left(\frac{\nu_i}{\nu_0}\right)^{\beta}\left(\frac{\nu_j}{\nu_0}\right)^{\beta}
\end{equation}
The amplitude $A$ and slope $\beta\approx -2$ has been obtained by
extrapolation from the fluxes and spectra of $208$ resolved point
sources in the WMAP-1~\cite{Bennett1} and $300$ point sources in
WMAP-3~\cite{hin_wmap06}.

The contribution from each region $\alpha =1,2,\ldots, 9$ to the
expected residual from unresolved point sources in our cross-power
spectrum of cleaned maps ${\mathbf i}$ and ${\mathbf j}$ which are
each linear combinations of $4$ DA's at different frequencies
(eq.~\ref{c_map}) is given by
\begin{equation}
C^{(PS) \alpha \mathbf{ij}}_l
=\sum_{i=1}^4
\sum_{j=1}^4 w^{i}_{l}({\mathbf i}) w^{j}_{l}({\mathbf j})
A\left (\frac{\nu_i}{{\nu_0}}\right)^{-2}\left
(\frac{\nu_j}{{\nu_0}}\right)^{-2},
\end{equation}
where $w^{i}_{l'}({\mathbf i})$ denotes the weight vector for the
cleaned map ${\mathbf i}$ and $\nu_0$ is a fiducial frequency. The
index pair $\mathbf{(ij)}$ are effectively a single index limited to the
$24$ `allowed' cross-spectra.  The total contribution is added up as
\begin{equation}
C^{(PS)}_l =
(1/24)\sum_{\mathbf{(ij)}=1}^{24}\sum_{\alpha}\sum_{l'}M^{(\alpha)}_{ll'}
C^{(PS)\alpha \mathbf{ij}}_{l'}
\end{equation}
where $M^\alpha_{ll'}$ the coupling matrix for the individual regions
where $\alpha = 1, 2 , ..., 9 $.

\begin{figure}
\begin{center}
\includegraphics[width=6cm,angle=-90]{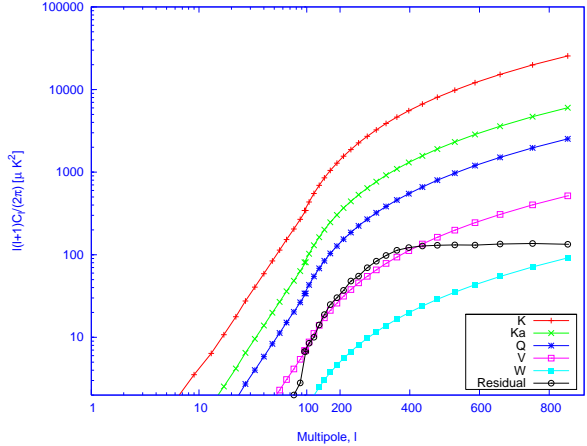}
\caption{Comparison of the residual power in WMAP due to unresolved
point source contamination (black) in the final averaged cross power
spectra in our method with the actual level point source contamination
in each the $5$ frequency channels. Note that the level of residual
contamination is in between the cleanest (highest) two frequency
channels (V \& W). More importantly, the residual point source
correction is constant with multipoles whereas the actual level of
contamination increases $\propto l^2$ with multipoles. This implies
that our method is efficient in canceling out the point source
contamination at high resolution.}
\label{ps_plus_5_channel}
\end{center}
\end{figure}

We use the above expression to estimate the unresolved residual point
source contamination in the final cross power spectrum.  We note in
passing that this estimation is entirely based upon the assumption
that the residual unresolved point source contamination is
statistically isotropic over the sky.  The residual power from
unresolved point sources in WMAP-1 {\it[WMAP-3]} is a constant offset
of $\sim 140~~\mathit{[100]} \mu K^2 $ for $l \gsim 400$ (and
negligible at lower $l$).  As seem in figure~\ref{ps_plus_5_channel},
this residual is much less than actual point source contamination in
Q, KA or K band and intermediate between V and W band point source
contamination. It is noteworthy that the method significantly tempers
the point source residual at large $l$ that otherwise is $\propto l^2$
in each map.

\subsection{Error Estimate on the Power Spectrum}
\label{Errorbars}
The errors on the final power spectrum are computed from Monte Carlo
simulations of CMB maps for every DA each with a realization of the
WMAP noise and common signal and diffuse foreground contamination.
For WMAP-1, we use the $110$ random noise maps made available at the
LAMBDA data archive.  For WMAP-3 year analysis random noise map
corresponding to each DA are generated by first sampling a Gaussian
distribution with unit variance. In the final step we multiply each
Gaussian variable by the number $\sigma_0/\sqrt{N_p}$ to form
realistic noise maps. Here $\sigma_0$ is the noise per observation for
the DA under consideration and $\sqrt{N_p}$ is the effective number of
observations at each pixel. We use the publicly available Planck Sky
Model to simulate the contamination from the diffuse galactic
(synchrotron, thermal dust and free-free) emission at the WMAP
frequencies. The CMB maps were smoothed by the beam function
appropriate for each WMAP's detector. The set of realistic DA maps
with noise, foreground and CMB are then passed through the cleaning
pipeline.  The common CMB signal in all the maps was based on a
realization of the WMAP `power law' best fit $\Lambda$-CDM
model~\cite{Spergel}.  Averaging over the power spectra from the
simulations we recover the model power spectrum, but for a hint of
bias towards lower values in the low $l$ moments. For $l=2$ and $l=3$
the bias is $-27.4\%$ and $-13.8\%$ respectively. However this bias
become negligible at higher $l$, e.g. at $l=22$, it is only $-0.8\%$.

The standard deviation obtained from the diagonal elements of the
covariance matrix is used as the error bars on the $C_l$'s obtained
from the data~\footnote{The beam uncertainty is not included here, but is
deferred to future work where we also incorporate non-circular beam
corrections~\cite{mit04}.}.  The covariance matrix of the binned power
spectrum is largely diagonal.

\section{Conclusion}
 
The rapid improvement in the sensitivity and resolution of the CMB
experiments has posed increasingly stringent requirements on the level
of separation and removal of the foreground contaminants.  Standard
approaches to foreground removal, usually attempt to incorporate the
extra information about the amplitude, spatial structure and
distribution of foregrounds available at other frequencies, in
constructing a foreground template at the frequencies of the CMB
measurements. These approaches could be susceptible to uncertainties
and inadequacies of modeling involved in extrapolating from the
frequency of observation to CMB observations.

We carry out an estimation of the CMB power spectrum from the WMAP
data that is independent of foreground model and evades these
uncertainties. The novelty is to make clean maps from the difference
assemblies and exploit the lack of noise correlation between the
independent channels to eliminate noise bias.  {\em This is the first
demonstration that the angular power spectrum of CMB anisotropy can be
reliably estimated with precision solely from the WMAP data
(difference assembly maps) without recourse to any external data.} Our
work is a clear demonstration that the blind approach to foreground
cleaning is comparable in efficiency to that from template fitting
methods and certainly adequate for a reliable estimation of the
angular power spectrum. Moreover, this method enjoys the advantage of
eliminating the uncertainties of modeling required to create the
foreground templates that estimate the contamination at the CMB
dominated frequencies by extrapolating from the observed emission flux
at very different frequencies where the foregrounds dominate.

Further, the tempered, flat, large $l$ behavior of residual from
unresolved point sources in this method holds promise for experiment
with finer angular resolution.  The understanding of polarized
foreground contamination in CMB polarization maps is rather
scarce. Hence modeling uncertainties could dominate the systematics
error budget of conventional foreground cleaning.  The blind approach
extended to estimating polarization spectra after cleaning CMB
polarization maps could prove to be particularly advantageous.

\section*{Acknowledgment}

The analysis pipeline as well as the entire simulation pipeline is
based on primitives from the Healpix package~\footnote {The Healpix
distribution~\cite{healpix} is publicly available from the website
http://www.eso.org/science/healpix.}. We acknowledge the use of
version 1.1 of the Planck reference sky model, prepared by the members
of Working Group 2 and available at www.planck.fr/heading79.html. The
entire analysis procedure was carried out on the IUCAA HPC
facility. RS thanks IUCAA for hosting his visits. We thank the WMAP
team for producing excellent quality CMB maps and making them publicly
available. We thank Amir Hajian, Subharthi Ray and Sanjit Mitra in
IUCAA for helpful discussions. We are grateful to Lyman Page, Olivier
Dore, Francois Bouchet, Simon Prunet, Charles Lawrence, Kris Gorski
and Max Tegmark for thoughtful comments and suggestions on this work.


\begin{thebibliography}{99}
\frenchspacing
%%%%%%%%%%%%%%%%
\bibitem{bouc_gisp99} F. R. Bouchet and
R. Gispert, New Astronomy {\bf 4},443, (1999).

\bibitem{Tegmark96} M. Tegmark, G. Efstathiou 
\mnras, {\bf 281}, 1297, (1996).

\bibitem{FIRAS} J. Mather {\em et al.},
\apj, {\bf 420}, 439, (1994); {\em ibid.}  {\bf 512}, 511 (1999).

\bibitem{Dodelson} S. Dodelson,  \apj, {\bf 482}, 577 (1997).

\bibitem{Tegmark2000a} M. Tegmark, D.~J. Eisenstein, W. Hu, \&
A. de-Oliveira-Costa, \apj, {\bf 530}, 133, (2000).

\bibitem{Bennett1} C.~L. Bennett {\em et al.}, \apjs, {\bf 148}, 97
(2003).

\bibitem{Tegmark98} M. Tegmark, \apj {\bf 502} 1, (1998).

\bibitem{hob98} M. P. Hobson {\em et al.}, \mnras, {\bf 300} 1 (1998).

\bibitem{main0203} D. Maino {\em et al.},  \mnras, {\bf 334}, 53
(2002); {\em ibid}, {\bf 344}, 544 (2003).

\bibitem{erik06} H. K. Eriksen et al., \apj {\bf 641}, 665, (2006). 


\bibitem{sah06} R. Saha, P. Jain \& T. Souradeep, Astrophys. J. Lett.,
{\bf 645}, L89, (2006).

\bibitem{hin_wmap03} G. Hinshaw {\em et al.}, \apjs, {\bf 148}, 135 (2003).

\bibitem{Tegmark} M. Tegmark, A. de Oliveira-Costa \& A. Hamilton,
Phys. Rev. D68, 123523 (2003).

\bibitem{Jarosik} N.Jarosik {\it et al.},  \apjs {\bf 145}, 413
(2003).

\bibitem{Bennett} C. ~L.Bennett {\em et al.}, \apjs {\bf 148}, 1
(2003).


\bibitem{Bennett2} C. ~L.Bennett {\em et al.}, \apj, {\bf 583}, 1
(2003).

\bibitem{WMAP} M. Limon, {\em et al.}, Wilkinson
Microwave Anisotropy Probe (WMAP): Explanatory Supplement, version
1.0, at the LAMBDA website.

\bibitem{Hinshaw1} G. Hinshaw {\em et al.} \apjs, {\bf 148}, 63 (2003).

\bibitem{Pablo} P. Fosalba \&  I. Szapudi, \apj, {\bf 617}, 95 (2004).

\bibitem{Patanchon} G. Patanchon, J.~F. Cardoso, J. Delabrouille \&
 P. Vielva, {\it preprint} [arXiv: astro-ph/0410280].


\bibitem{mit04} S. Mitra, A.~S.Sengupta and T. Souradeep,
Phys.~Rev.~D. \textbf{70} 103002 (2004).

\bibitem{irv_noncirc} T. Souradeep, S.Mitra, A.~S. Sengupta~, S. Ray
 and R. Saha , {\it (in these proceedings)}.

\bibitem{Hivon} E. Hivon {\em et al.}, \apj {\bf 567}, 2 (2002).

\bibitem{knox01} L. Knox, N. Christensen \& C. Skordis, \apj {\bf
563}, L95, (2001).

\bibitem{jew04} J. Jewell, S. Levin \& C.H. Anderson, \apj {\bf 609},
1, (2004).

\bibitem{wan04} B. D. Wandelt, D. L. Larson \& A. Lakshminarayanan, 
Phys.~Rev.~D. {\bf 70}, 083511 (2004).

\bibitem{Saha} R. Saha, P. Jain,  \&
 T. Souradeep, {\it in preparation}.

\bibitem{Amir1} A. Hajian \& T. Souradeep, \apj, {\bf 597}, 5 (2003).

\bibitem{Amir2} A. Hajian {\em et al.} \apj, {\bf 618}, 63 (2004).

\bibitem{Amir3} A. Hajian, T. Souradeep, {\it preprint} [arXiv:
astro-ph/0501001].

\bibitem{irv_stataniso} S. Basak, A. Hajian \& T. Souradeep, {\it (in
these proceedings)}, [arXiv: astro-ph/0607577]

\bibitem{hin_wmap06} G. Hinshaw {\em et al.}, {\it preprint} [arXiv:
astro-ph/0603451].


\bibitem{wmap3reanal} H. K. K. Eriksen {\em et al.}, {\it preprint}
[arXiv:astro-ph/0606088].

\bibitem{Spergel} D.  Spergel {\em et al.}, Astrophys. J. Suppl., {\bf
148}, 175, (2003); {\it ibid} {\it preprint} [arXiv: astro-ph/0603449].

\bibitem{healpix} K.~M. Gorski {\em et al.}, [arXiv:astro-ph/9905275];
Gorski, K.~M. {\em et al.}  [arXiv: astro-ph/9812350].


\end{thebibliography}
\end{document}